\documentclass{vgtc}                          

\ifpdf
  \pdfoutput=1\relax                   
  \usepackage{graphicx}                
  \DeclareGraphicsExtensions{.pdf,.png,.jpg,.jpeg, .svg} 
\else
\usepackage{svg}
  \usepackage{graphicx}                
  \DeclareGraphicsExtensions{.eps}     
\fi%
\usepackage{enumitem}
\usepackage{hyperref}

\graphicspath{{figures/}{pictures/}{images/}{./}} 
\usepackage{comment}
\usepackage{microtype}                 
\PassOptionsToPackage{warn}{textcomp}  
\usepackage{textcomp}                  
\usepackage{mathptmx}                  
\usepackage{times}                     
\usepackage{cite}                      

\onlineid{1017}

\vgtccategory{Research}

\vgtcinsertpkg

\title{ESID: Exploring the Design and Development of a \\ Visual Analytics Tool for Epidemiological Emergencies}

\author{Pawandeep Kaur Betz
\thanks{These authors contributed equally to the study.}
\thanks{e-mail: pawandeep.kaur-betz@dlr.de}
\\ %
      \parbox{1.4in}{  \scriptsize  \centering German Aerospace Center}
\and Julien Stoll\footnotemark[1] \\ %
    \parbox{1.4in}{ \scriptsize \centering Hochschule für Gestaltung Schwäbisch Gmünd} %
\and Valerie Grappendorf\footnotemark[1]\\ %
    \parbox{1.4in}{ \scriptsize \centering Hochschule für Gestaltung Schwäbisch Gmünd} %
  \and Jonas Gilg \\ %
     \parbox{1.4in}{\scriptsize  \centering German Aerospace Center}
\\
\and Moritz Zeumer\\ %
    \parbox{1.4in} {\scriptsize \centering German Aerospace Center}
\and Margrit Klitz \\ %
     \parbox{1.4in}{\scriptsize \centering \scriptsize  German Center for Neurodegenerative Diseases}
\and Luca Spataro\\ %
     \parbox{1.4in}{\scriptsize \centering German Aerospace Center}
  \and Anna Klein \\ %
     \parbox{1.4in}{\scriptsize  \centering Helmholtz Centre for Infection Research}
\\
\and Lena Rothenhaeusler\\ %
     \parbox{1.4in}{\scriptsize  \centering Helmholtz Centre for Infection Research}
\and Hartmut Bohnacker \\ %
    \parbox{1.4in}{\scriptsize  \centering Hochschule für Gestaltung Schwäbisch Gmünd}
\and Hans Kraemer \\ %
    \parbox{1.4in}{\scriptsize  \centering  Hochschule für Gestaltung Schwäbisch Gmünd}
  \and Michael Meyer-Hermann
  \thanks{Principal investigators of this study.}\\ %
    \parbox{1.4in}{ \scriptsize \centering Helmholtz Centre for Infection Research \\}
\\
\and Sybille Somogyi\footnotemark[3]\\ %
     \parbox{1.6in}{\scriptsize  \centering  Academy for Public Health Services}
\and Andreas Gerndt\footnotemark[3] \\ %
  \parbox{2.0in}{\scriptsize \centering German Aerospace Center and \\ \scriptsize  University of Bremen}    
\and Martin J. Kühn 
\footnotemark[3]
\thanks{e-mail: martin.kuehn@dlr.de}\\ %
     \parbox{1.4in}{ \scriptsize \centering German Aerospace Center  }  
     } %

\abstract{Visual analytics tools can help illustrate the spread of infectious diseases and enable informed decisions on epidemiological and public health issues. To create visualisation tools that are intuitive, easy to use, and effective in communicating information, continued research and development focusing on user-centric and methodological design models is extremely important. 
As a contribution to this topic, this paper presents the design and development process of the visual analytics application 
ESID (\textit{Epidemiological Scenarios for Infectious Diseases}). ESID is a visual analytics tool aimed at projecting the future developments of infectious disease spread using reported and simulated data based on sound mathematical-epidemiological models. The development process involved a collaborative and participatory design approach with project partners from diverse scientific fields. 
The findings from these studies, along with the guidelines derived from them, played a pivotal role in shaping the visualisation tool.

} 

\CCScatlist{
  \CCScatTwelve{Human-centered computing}{Visu\-al\-isa\-tion}{User Studies}{}
}

\begin{document}

\firstsection{Introduction}
\maketitle

Visual analytics (VA) tools are increasingly used in the field of epidemiology and public health~\cite{carroll2014visualization, preim2020survey}, particularly, for the surveillance and management of infectious diseases with the potential of emerging pandemics. Scientists need to analyse a large amount of data at different scales for predictions with mathematical models. This is where VA tools help in gaining insights from the large and complex data sets. Based on these insights, scientists can advise decision makers involved in the control of infection events to make informed decisions.

Developing VA tools for epidemiology and public health necessitates collaboration among experts from diverse fields. Trans-disciplinary~\cite{Hadorn2008handbook} and participatory design methodologies~\cite{janicke2020participatory} are crucial in ensuring that these tools effectively address the specific requirements and needs of various user groups.

In acute emergency situations, the rapid development of visual analytics tools ~\cite{dixit2020rapid, afzal2020visual, chen2022rampvis} becomes a priority to address immediate needs. 
The constraints of such scenarios can make it challenging to fully implement participatory design studies that require long-term planning and extensive interactions with project partners. However, even in time-constrained situations, incorporating established guidelines, usability principles derived from previous design studies, and insights from literature analysis can facilitate the creation of user-centric approaches.

The contribution of the current paper is the presentation of a detailed design and development process of our visual analytics tool ESID (\textit{Epidemiological Scenarios for Infectious Diseases }). This tool is the result of our long-term study to understand diverse user requirements. Furthermore, we also contribute by recommending important guidelines for the development of visual analytics tools for epidemiology and public health. These guidelines are derived from our design study, our extensive experience in designing ESID and understanding the specific requirements of the domain. By sharing these guidelines, we aim to provide valuable insights to researchers and practitioners in the field, helping them develop effective and user-centric visual analytics tools. 

ESID development is still in progress, and new features are updated based on feedback from all the stakeholders and our ongoing studies. Once publicly available, ESID will serve as a public health dashboard for policymakers, healthcare professionals, and the general public, which provide disease related information for real and simulated data. ESID is published under a permissive open source Apache 2.0 license. The repository is available at \url{https://github.com/DLR-SC/ESID}. 
It is citable as \url{10.5281/zenodo.7681429}.

Currently it is deployed only for analysis of SARS-CoV-2 cases, however, it is designed and developed to explore and visualise the spread of any type of infectious disease. SARS-CoV-2 served as a use case for the design study and the development of this tool.
\section{Related work}
\label{sec:related_work}

A significant number of studies have been focusing on visualising the spread of COVID-19. Various visualisation techniques have been employed to communicate the complex patterns and trends associated with the pandemic. Heatmaps and Choropleth Maps~\cite{li2021visualizing} are used to represent the intensity of COVID-19 cases across geographical regions. Time-Series Charts, used to display the progression of COVID-19 cases, hospitalisations or deaths over time, are commonly used to understand the temporal dynamics~\cite{cai2021temporal, neelon2021spatial} of the pandemic. Network graphs~\cite{kapoor2020examining} are used or modelled to represent the spread of COVID-19 through connections between individuals or locations. While these studies have made valuable contributions to the field, it is important to note that their focus is often limited to a particular aspect of insight related to a specific visualisation type. Each visualisation technique offers unique advantages in conveying specific information, but no single technique can capture the full complexity of the pandemic. Therefore, a tool that provides a combination of different visualisation approaches and techniques is necessary for comprehensive understanding of the disease.

Many different visualisation techniques are grouped together during the creation of data or visualisation dashboards. Disease related data dashboards show multidimensional insights into disease patterns and often allow users to explore different regions, zoom in on specific areas, visualise temporal variations via charts, and other visual elements. They are commonly known as Public Health Dashboards (PHDs) ~\cite{preim2020survey} and have been popularly used by federal health organisations and research institutions. For example, there are dashboards from
Johns Hopkins University Center for Systems Science and Engineering (\url{https://coronavirus.jhu.edu/map.html})\cite{dong2020johnhopkin}, World Health Organization (WHO) (\url{https://covid19.who.int/}), European Centre for Disease Prevention and Control (ECDC) (\url{https://www.ecdc.europa.eu/en/covid-19/country-overviews}),
to only name a few. These dashboards are, however, limited to the task of data reporting and do not provide forecasting of the future trend of a disease. It should be noted that there are also plenty of studies, e.g.,~\cite{contreras2021challenges, quilty2021quarantine, afzal2020visual, dixit2020rapid, chen2022rampvis}, that have been providing forecasts on COVID-19. However, due to the urgent needs of the pandemic and the background of main developers in mathematics and epidemiology, most of them did not undergo a full user-centric design process. Nevertheless, they were driven by scientific excellence and the desire to contribute.

Considering all above aspects, ESID application advances the science of VA applications for epidemiological analysis: 1) By providing an epidemiological visual analytics tool based on a carefully curated user-centric design process. 2) A comprehensive tool that combines different visualisation approaches and uses the full potential of different techniques. 3) By leveraging historical data and applying predictive models, the application provides insights into future disease trends, allowing users to make more informed decisions.


\section{Design Process}
\label{sec:design_process}
 
In this section, we describe the design process of ESID, incorporating continuous feedback from stakeholders and researchers at all steps.  
ESID contributes to different long-term scientific projects, whose common goal is to develop a software system for the surveillance, control and prevention of epidemic outbreaks. The development of this software is done with the close collaboration of experts from different domains which we categorise as follows:  

\begin{enumerate}
    \item \textbf{Core team:} The core team is responsible in the creation of the software product. Based on their expertise, we divide them into the following domains:
    \begin{itemize}
        \item Epidemiology, Public Health, Immunology and others: Expertise in infectious diseases dynamics and mechanisms in disease transmission.
        \item  Mathematics, Scientific Computing and others: Expertise in leveraging mathematical models to the large scale and efficient treatment of large data sets.
        \item Design, UX development: Carrying the requirement analysis and design study to design a frontend.
        \item Visualisation Scientists: Expertise in the creation of visualisation software and performing usability testing.
    \end{itemize}
    
    \item \textbf{Involved stakeholders:} This team provides primary data and feedback on simulation, modelling and interface design. Team members come from epidemiology or public health research institutions, local health authorities, or are external scientists from domains such as epidemiology, immunology, virology, or data science.

    \item \textbf{Interested stakeholders:} This group is interested in the output of the results and includes, for example, policymakers and political institutions, health facilities, educational facilities, companies and interested citizens.
   
\end{enumerate}

In our design study, we first investigated the needs and challenges of our target user groups during the spread of COVID-19. Our idea that a tool that assists public health and policymakers in understanding the spread of the virus visually had been confirmed to be a valuable contribution. We also reviewed existing COVID-19 visualisation tools and scientific literature. We tried to depict the process of visualisation design via notion of double diamond design process model~\cite{council2007eleven, neelon2021spatial}. In this process model, double diamonds represent exploring an issue more widely or deeply and then taking focused action. We designed the software concept following their four suggested phases of discover, define, develop and deliver. Based on our study, we have divided these suggested phases into eight steps. 

 \begin{figure}[!htbp]
    \centering
    \includegraphics[width= \linewidth]{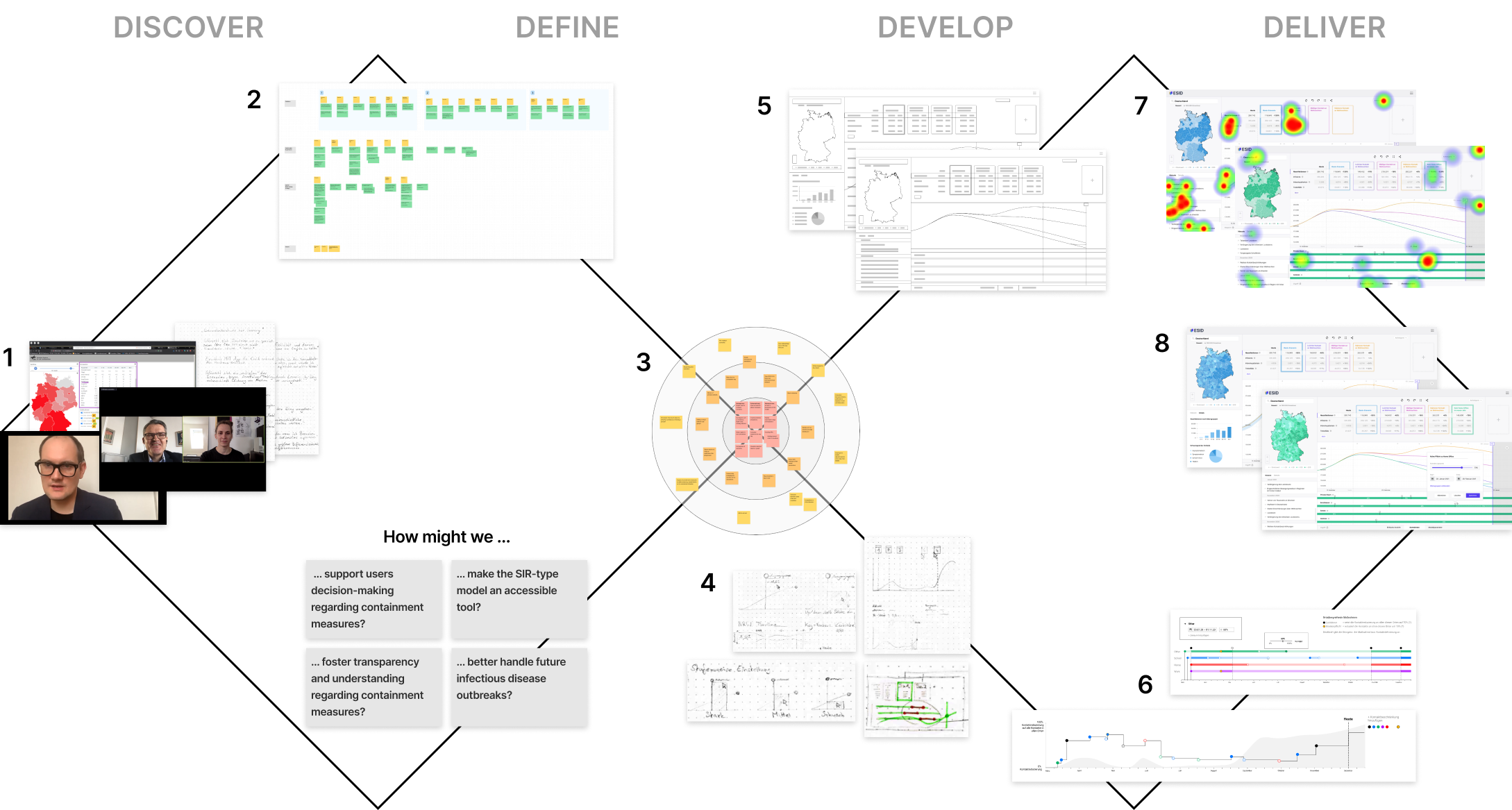}
    \caption{\textbf{Design process of the ESID visual analytics tool.} Based on our study, we have divided this process into eight steps: 1) Requirement gathering, 2) Synthesising, 3) Prioritisation, 4) Ideation, 5) Wireframing, 6) Prototyping, 7) Evaluation and 8) Visual Design. High resolution images are available in supplementary material 1, 2and 3.}
    \label{fig:design process}
\end{figure}

In the \textit{Discover} phase, we gathered the requirements from our target group. We conducted interviews with 13 target users. During the interviews, notes were taken, and whenever possible, the interviews were also recorded for later analysis. This qualitative research aimed to understand the interviewees' challenges and needs in understanding the future dynamics of the COVID-19 pandemic. To gain additional information, we analysed passive information via published speeches from political advisers, experts in the public health sector and surveys from different verified sources. Then we \textit{synthesised} the notes and insights of the interviews and grouped the recurring themes in clusters.  We derived different features from these themes. Then during step three, we \textit{prioritised} these features pragmatically based on the visualisation mantra of overview first and then details on demand~\cite{schneiderman1996the}. In overview, all features that are associated in depicting the spatio-temporal pattern of the virus spread were considered. Then for details, we considered features, where interactive components are needed to filter more information from the data. These features receives lower priority and thus moved to the outer circle (see supplementary material 2).

\begin{figure}[!htbp]
  \includegraphics [scale=0.4]
  {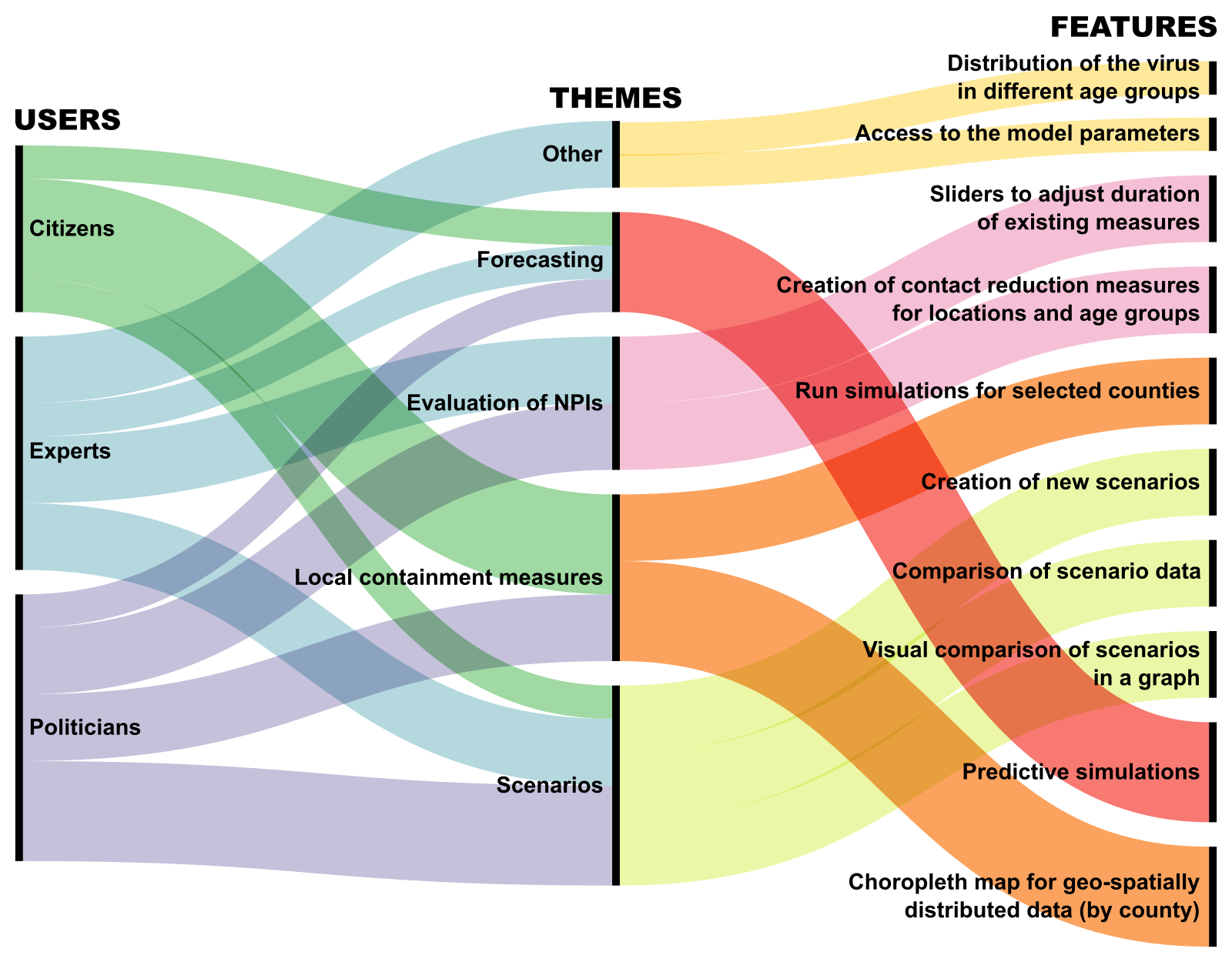}
    \caption{ \textbf{Themes and features emerged from the study.} This alluvial diagram shows 
 the common themes emerged from the users during the requirement gathering phase. It also shows the derived features for the development of the visual analytics applications. Different visual components that are derived from these features are presented in \autoref{tab:development}. }
    \label{fig:requirement_analysis}
\end{figure}

The mapping of themes to their features is shown in \autoref{fig:requirement_analysis}. This analysis highlighted overlap for certain themes across the three user groups. Our research led to new insights and confirmed initial assumptions that validated the need for a publicly accessible simulation tool. 

As shown in \autoref{fig:requirement_analysis}, members of all user groups would like to know the results based on the current development of the infectious disease dynamics (\textbf{Forecasting}). They would also like to get predictive results based on different intervention measures (\textbf{Scenarios}). Ingredients of these scenarios are: face mask mandates, testing and tracing strategies, venue closures, work from home, as well as vaccinations if available. As one politician said, \textit{"It would be exciting to see the impact of taking and not taking different actions."} In order to come to informed decisions, users would best like to know the impact of interventions. Therefore, they would also like to have the flexibility to create new scenarios and then visually compare the outcomes with other scenarios. To accomplish these requirements: scenario cards and configurable scenario filters are being developed. 
 
The need for an \textbf{Evaluation of NPIs} (NPIs: non-pharmaceutical interventions) 
to reduce the workload on health officials and medical staff is very closely coupled to the \textbf{Scenarios}. In addition to the visual comparison of scenarios, features are needed to adjust the length or strength of interventions such as mask mandates or venue closures. Also as NPIs are often targeted to particular subgroups of the population such as pupils, workers or elderly, we  provide a configurable filter based on age groups, or other available sociodemographic properties. In supplementary material 4, we show the design of this component in our application.

Another common theme among the interviewees was the spread of the pandemic and the effect of different containment measures in their own regions or counties~\textbf{(Local containment measures)}. One epidemiologist said, \textit{"If we notice that locally, and I emphasise locally, that an
outbreak occurs, then we have to intervene very massively and locally. That has a great advantage: the local environment at other places can continue economic performance and normal life."} To fulfil this need, we decided to include a county/district-based choropleth map.
Apart from that, interviewees also asked for a configurable and flexible tool wherein the model parameters can be easily changed, and the results can be seen without much delay and effort. In supplementary material 5, we show the design of this component, which is under development.

\begin{figure*}[!htbp]
    \centering\includegraphics[width= .9\linewidth]{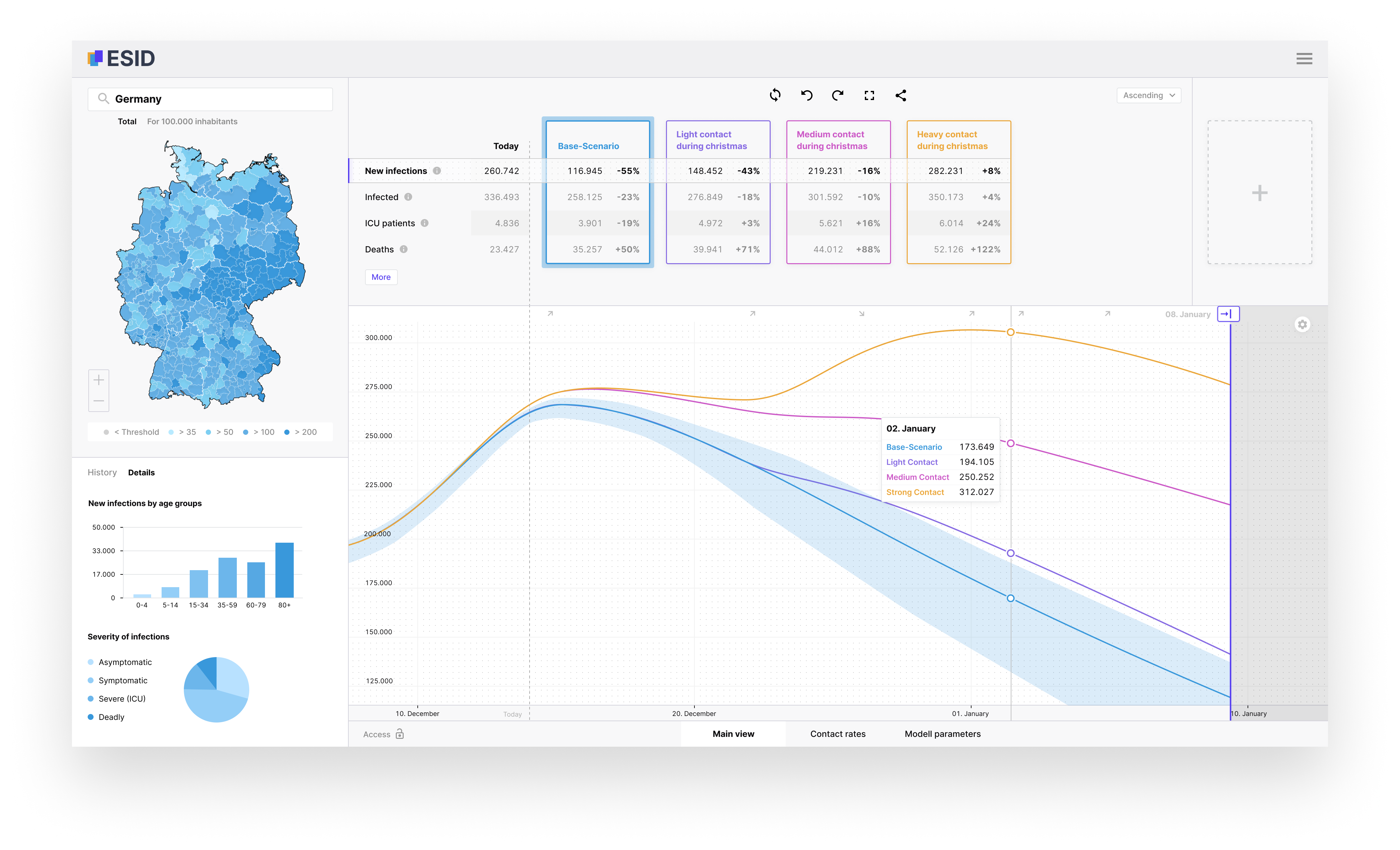}
    \caption{Frontend of the ESID visual analytics application}
    \label{fig:ESID}
\end{figure*}

The interface design was started after the \textit{prioritisation} stage (see \autoref{fig:design process}), wherein we concretely prioritised the needs and features for further development. We started with the process of \textit{Ideation} (\autoref{fig:design process}, step 4) in which we used sketches to capture and communicate first ideas. After a few iterations, these served as the basis for creating \textit{wireframes} (\autoref{fig:design process}, step 5). Through different wireframes, we ensured that all necessary features were integrated into the information architecture and layout of the user interface. These wireframes were then converted into \textit{prototypes} (\autoref{fig:design process}, step 6) to test them with the target users. The prototypes were adjusted in an iterative fashion based on user \textit{evaluation} (see \autoref{fig:ESID}, step 7 ). From these learnings, the \textit{visual design} (see \autoref{fig:ESID}, step 8 ), was created which is being further developed as a frontend of the ESID visual analytics application.

\subsection*{Guidelines for a visual analytics infection control tool} \label{sec:guidelines}

In the following, we present the guidelines derived from our design study, our experience from the design phase of ESID, and understanding of the specific requirements of the domain. We hope that these points can serve as essential guidelines for designing and developing visual analytics applications for epidemiological analysis. 

\begin{enumerate}
    \item \textbf{Reliable and empirical data}\label{1}: 
    The reliability and accuracy of predictions in disease analysis heavily relies on using reliable and empirical data from trusted sources; like national and international institutions in Public Health, official surveillance systems, sound surveys, and accredited research institutions. These institutions follow rigorous protocols and adhere to strict regulations for data collection and analysis. As a result, their data is more likely to be accurate and standardised, which can further minimise the risk of false predictions and rumours.

    \item \textbf{Transparency in communicating data}:  In visualising predicted information, visual analytical tools need to make sure that there is a transparency in communicating a) uncertainty in the data and results, and b) sensitivity in the results. \begin{itemize}
        \item \textbf{Uncertainty}: \label{2.1} Uncertainty is inherent in simulations, especially during outbreaks or emergencies, where data and information may be rapidly evolving or incomplete. Lack of this feature in a software could lead to (retrospective) misinterpretation of data and hinder trust among citizen and decision makers. 
        
        \item \textbf{Sensitivity of predicted results}: \label{2.2} From analysing the results of our interviews, we realised that politicians understand the fact that models are based on assumptions. Thus it is important to provide tools or interactive ways to visualise the effect of different factors influencing these assumptions.
\end{itemize}

\item \textbf{Different viewing levels}: \label{3} The dynamics of infectious diseases follow complex patterns and its predictions comes at a certain cost. According to two reasons, we decided for different access levels.
\begin{itemize}
    \item A non-scientific user may be overwhelmed with providing estimates for model parameters. On the other hand, experts want to understand how results have been obtained. Thus, simple and expert views should be available.
    \item A number of meaningful scenarios or results can be precomputed (e.g. overnight) and provided to all users. However, even if optimized and quick-to-evaluate models as from machine learning contexts are used, demanded results for new settings may not be available in seconds or come at a high cost due to the use of supercomputers. In order to resolve this issue different user levels can be introduced and, e.g., only admin level users may trigger model runs on supercomputers. 
\end{itemize}
    
     \item \textbf{Spatial overview}:\label{4} Visual analytics tools should provide views of the dynamics at different spatial levels. While at country level, users may want to see comparisons with other nations, at county and regional level, users may be interested in much finer granularity about outbreaks and hotspots.

     \item \textbf{Comparative analysis}: \label{5}A Visual analytics tools for epidemiological emergencies should include different interactive techniques and filters to do comparisons at different dimensions: spatial, temporal, demographics, or infection states and see the results in an intuitive way via the use of visualisation types. These comparative visualisations would also be an important way to provide transparency in communicating the data. 
     
     \item \textbf{Detection and warning}: \label{6}A useful tool for emergencies should also provide early warning for any prospective outbreak or hotspot formations. By the use of different chart types and visual cues, this information needs to be effectively presented on the application.

     \item \textbf{Visualising mobility and contact patterns}:\label{7} By visualising mobility data and (assumed) contact patterns, policymakers and researchers can gain valuable insights into population movements and behaviours, which are crucial factors in the transmission dynamics of a pandemic. Thus it contributes in enabling evidence-based decision-making and targeted interventions to control and mitigate the impact of the pandemic.
     
     \item \textbf{Flexibility, Compatibility and Reusability}:\label{8} In order to allow a good user experience, the following points need to be considered.
      \begin{itemize}
          \item Interactive and configurable: The tool should offer interactive features that allow users to explore and manipulate the visualisations based on their specific needs.
          \item Immediate reactions: In order to allow interactivity, the tool needs to respond immediately on user actions. For time-consuming backend or simulation tasks, solutions need to be found.
          \item User-friendly interface: The tool should prioritise an intuitive and user-friendly interface, ensuring that users can easily navigate and interact with the visualisations without requiring extensive technical knowledge. 
          \item Customisable: Offering customisation options is beneficial for users who have specific requirements or want to tailor the tool to their unique contexts. This can include options to choose different map projections, colour schemes etc. 
          
         \item Platform-independence: The tool should be platform-independent to allow all interested users to use it. 
         \item Low access barriers: As software installation often is not allowed for many users, a user friendly web page solution is advantageous.
         \item Integration with third-party solutions: The tool should be designed in a way that allows it to be easily integrated into third-party platforms or solutions. 
         \item Dependencies: Dependencies should be chosen with care such that discontinuation of a dependency does not critically affect the use of the VA tool.
         \item Extensibility: As infectious disease spread is also dynamic with respect to mitigation options through, e.g., newly available vaccines it should be foreseen that provided information or data need to be changed or extended. 
        \item Reusability: The tool and the model needs to be open source and available for others to replicate. By allowing other scientists build upon it, the tool fosters transparency and promotes further research and development in the field.     
         \end{itemize}
     
     \item \textbf{Little room for interpretation}:\label{9} Addressing concerns about misinterpretation and misuse of simulations is crucial when developing dashboards for infectious disease dynamics. The majority of our users from politics, public health system and science sector had indicated this fear. As a solution to that,  features like 1) user management for controlled access and 2) verifiable scenarios with descriptive text features can help mitigate these risks. 
  
\end{enumerate}
Dashboards for disease analysis are difficult to evaluate because of the diversity of users and use cases~\cite{ansari2022development}. Moreover, it can be particularly difficult to conduct user studies and gather requirements during rapid development of software or when direct access to users is limited. In such scenarios, guidelines, usability checklists, heuristics and principles  derived from other research and existing knowledge could become valuable tools for guiding the design and development process.
From our literature analysis, we realised that guideline no. \ref{1}, \ref{2.1}, \ref{5}, \ref{6} are also been discussed as user tasks in~\cite{preim2020survey} and that needs to be fulfilled by a visual analytics software for Public Heath. In their paper about future of public health dashboards, the authors of~\cite{dasgupta2022future} alerted on a need of dashboard based on audience segmentation. Wherein, they have asked for different views of the dashboards for different target audience. In relation to which our guideline \ref{9} and \ref{3} suggests user profiling and different access levels. Similarity of our proposed guidelines with established guidelines and best practices validates our results and recommendations. It also indicates that our guidelines address important considerations and requirements for designing effective disease analysis dashboards.

\section{ESID Visual Analytics Tool} \label{sec:ESID}
ESID is a visual analytics application -- developed to show the results of the simulations from a hybrid graph-equation-based model as introduced in~\cite{kuhn2021assessment}, but it should also be adaptable and extendable to other models and model types. The model can be stratified for different age groups and takes into account the properties of the infectious disease as well as human mobility and contact behaviour in the four locations home, school, work, and other. Since its first release in 2020, this model has been extended iteratively according to the developments of the disease, its variants and inclusion of the vaccination in 2021 and 2022; see~\cite{koslow2022appropriate} and~\cite{memilio}. Currently it provides results for 21 to 27 different infection states, basically, the states \textit{Susceptible}, \textit{Exposed}, \textit{Infectious non-symptomatic}, \textit{Infectious symptomatic}, \textit{Infected Severe}, \textit{Infectious Critical}, \textit{Dead} subdivided into three subpopulation with no or almost no immunity, people with partial and with highest immunity against Sars-CoV-2 infections. Furthermore, subdivision into detected and undetected commuters is possible. Immunity is gained via vaccinations or previous infections.

In its current form, ESID provides spatio-temporal exploration of data based on different infection states, forecasting based on different scenario definitions, comparisons of simulations results, filtering of data as well as options to download or export data and visualisations. It is developed as a single page web application based on the React framework (\url{https://reactjs.org/}) with Typescript (\url{https://www.typescriptlang.org/}). Further information on, e.g., mobility are planned to be shown in overlays or additional tabs. By the use of backend APIs, the results of several runs and differing parameters of our infection dynamics model are delivered. ESID reads a predefined data format and, therefore, could also visualise output from other models as long as they comply with the format definition. 

\autoref{fig:ESID} shows a mockup of ESID. On the left hand side is a map of the German counties with a customisable heat legend. At the top section are four cards representing four different simulation scenarios: blue, purple, pink and yellow.
Left of the scenario cards is a list of infection states or aggregations of infection states, like \emph{infected}, \emph{hospitalized}, and \emph{dead}. The bottom portion contains a line chart comparing the scenarios over a predefined timeline. This tool has multiple ways to interact, filter and select data. \autoref{tab:components} shows the relationship among different components and how they communicate with each other to visualise data. It can be read in the following way: The map shows the data of one scenario for one compartment on one date over all districts.
Since it shows all districts, it also serves as a selection tool for districts. The chart shows the data of all scenarios of one compartment over all days of one district. Since it shows all dates, it also serves as a selection tool for dates.

\begin{table} [!b]
	\begin{tabular}{|r| c| c| c| c|}\hline
		\textbf{} & \textbf{Scenario} & \textbf{Compartment} & \textbf{Date} & \textbf{District} \\\hline
		\textbf{Card} & select & all & 1 & 1 \\\hline
		\textbf{List} & none & select & 1 & 1 \\\hline
		\textbf{Chart} & all & 1 & select & 1 \\\hline
		\textbf{Map} & 1 & 1 & 1 & select \\\hline
	\end{tabular}
\\
\caption{Displayed information and interaction between visualisation components of ESID.}
\label{tab:components}
\end{table}

The user interface consists of three major components that provide interaction, changing the shared context, and visualising context-relevant data. These three components achieve most of the needed features derived from our requirement analysis process presented in \autoref{fig:requirement_analysis}. In \autoref{tab:development}, we show the role of each of these component. 

\begin{table}[b!]
\small
\begin{tabular}  { |l|p{5,8cm}| } 
     \hline
    Visual Components  & Derived Features 
    \\  \hline 
     Choropleth Map &
  \vspace{-0.75\baselineskip}
  \begin{itemize}[noitemsep, left=0pt, topsep=0pt]
        \item Choropleth map for geo-spatially distributed data (by county)
        \item Selectable counties to filter data in other components 
         \item A search bar with auto-complete function helps finding the county without geographical knowledge
  \vspace{-0.75\baselineskip}
  \end{itemize} 
    \\  \hline
 Scenario Cards &
  \vspace{-0.75\baselineskip}
 \begin{itemize}[noitemsep, left=0pt, topsep=0pt]
    \item Comparison of scenario data by infection states
    \item Stratification of the scenario data in different age groups
    \item Selectable infection states to filter data in other components
    \item Indicating increase or decrease compared to the start of the simulation    
    \item Trigger new simulation with customised parameters
  \vspace{-0.75\baselineskip}
\end{itemize} 
     \\  \hline
  Timeline &
  \vspace{-0.75\baselineskip}
  \begin{itemize} [noitemsep, left=0pt, topsep=0pt]
    \item Visual comparison of scenario data over time
    \item Temporal foresight through predictive simulations
    \item Sliders to adjust the duration of different measures for new simulations
  \vspace{-0.75\baselineskip}
\end{itemize}
\\  \hline
    \end{tabular}\\ 
    \caption{Role of the three major components of the ESID interface (see \autoref{fig:ESID}). The three components implement the derived features from our requirement analysis phase, as presented in \autoref{fig:requirement_analysis}.}
\label{tab:development}
\end{table}

\begin{figure}
    \centering
    \includegraphics[width=1\linewidth]{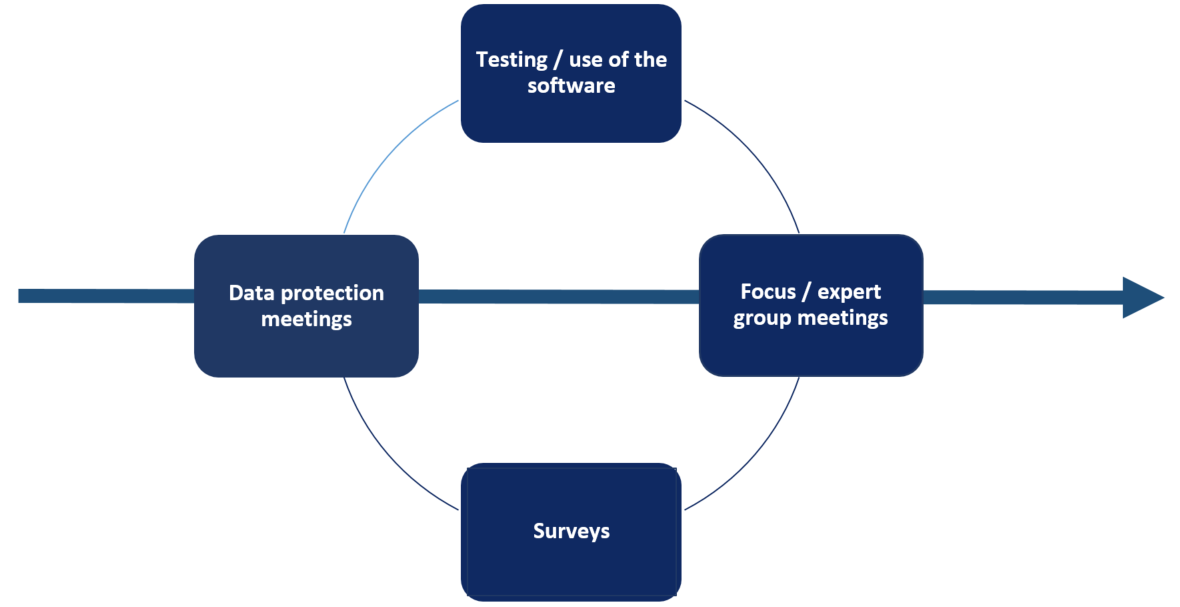}
    \caption{Further development of the application through user evaluation and feedback.
}
    \label{fig:user-feedback}
\end{figure}

To fulfil the guidelines provided in \autoref{sec:design_process}, ESID is based on the scenario outcomes supplied by a model created by several German research institutions and data is collected from the Robert-Koch Institute. To be transparent, we show uncertainty through coloured half-transparent areas on the line charts and through tool tips. Further studies are being conducted to know the most appropriate techniques to display uncertain information. To obtain a user-friendly interface, one of the main goals of point 8 of the guidelines (Flexibility, Compatibility and Reusability), we permanently consider the needs of the target users in local health authorities. 
For this purpose,  interviews had been conducted before and during the the design of the mockup and the development of a prototype. Based on a minimal viable product, interviews and surveys have been performed. These are furthermore performed regularly with the development of new features in order to match needs and functionality. During the entire project period user's opinions, evaluations and feedback will be integrated in the sprint-based development process. In paticular, regular expert meetings have been established. These monthly meetings with a handful of pilot health authorities are designed to have a personal exchange with the users and experts and to discuss relevant aspects. The goal of these meetings and regular surveys based on the user experience is to improve the functions of the ESID frontend. As a basis for the evaluation of the usability, the system usability scale of John Brooke \cite{brooke1996sus} was used as well as the customer satisfaction score and net promoter score. Our initial survey on the minimal viable product achieved an average score of 79,5\% (Good) (95\% CI 76,4;82,6). The system usability score was calculated for each participant. The overall optimisation of the ESID platform is an iterative process that will further be boosted through the involvement of health authorities; see \autoref{fig:user-feedback}.

Currently, we do not provide a feature for visual sensitivity analysis, however, we have planned this feature and its design would look as in supplementary material 5. The interactive choropleth map \autoref{fig:ESID} shows an overview of the country and for individual counties via mouse click. In the current version, the comparison can be made at temporal and spatial level via tool tips and filters. Through these filters, users can compare the results from different age groups and infection states. For instance, figure of group filter in supplementary material 4  depicts the effect of age group filter on the representation of the simulated results. 

\section{Discussion and Conclusion}
\label{sec:discussion}

The development of an interactive visual analytics tool for epidemiological emergencies requires the expertise from many domains. In our design process we tried to include experts from as many different domains as possible to satisfy the different and heterogeneous requirements. In addition, as stakeholders and participants from diverse communities potentially lack a common vocabulary, it has to be ensured that the final tool can be used intuitively from all parties. Furthermore, the situation, needs, and foci of interested stakeholders are dynamic and may vary with the course of an epi- or pandemic. Although we thoroughly tried to identify all these needs and map these to requirements and guidelines, our lists may be incomplete in specific situations or for individual stakeholders.

Certainly, the presentation is driven by the available (reported) data and, e.g., mobility patterns can only be shown if they can be identified reliably. So far, we have not yet designed the visualisation of reported individual contacts or transmission chains as this information was either non-existing, incomplete or unavailable to us due to data protection and other reasons. However, this information may be extremely important for certain stakeholders. We therefore acknowledge the fact that a change in available data may trigger a redesign of the guidelines or adaptation of the tool. Continuous feedback mechanisms and agile development and improvement is therefore important for the success of the VA tool.

Last but not least, our focus was on respiratory diseases with SARS-CoV-2 and COVID-19 as a use case. For other transmission ways, additional guidelines would be needed or some of our own would have to be adapted.

\subsection*{Conclusion}
The current paper reports on the design and development of ESID, a visual analytics tool for epidemiological emergencies. ESID is capable of showing the projections of mathematical models on future developments of infectious disease dynamics. The tool was developed through a collaborative and participatory design process with project partners from different scientific fields. The requirement gathering phase was instrumental in shaping our visualisation tool. It emphasized that policymakers would benefit from information on current and future trends of infectious diseases on different scales to dynamically decide on the implementation of effective interventions. The results of this phase are summarised in \autoref{fig:requirement_analysis} showing the primary user needs in the application. Based on this study, we have also presented important guidelines to consider in the design and development of a corresponding visual analytics tool.  

The goal of this study and in return this tool is to provide a platform (with its development guidelines) for quick and sound evaluations of the effect of different non-pharmaceutical interventions in order to facilitate evidence-based political decisions. To guarantee the usefulness of the tool, all requirements of the project partners and the users were carefully considered and integrated into the design. Ongoing iterative refinements based on user tests are in process to ensure good usability of the user interface and features that fulfil user needs. 
Further research is going on to effectively visualise the uncertainty and the mobility in the software. Furthermore, also the granularity and type of non-pharmaceutical intervention (sets) needs to be evaluated with the users. 

All in all, we tried to lay the ground for future tools that can help in mitigating the spread of infectious diseases using insights on a local scale. However, our work is still in the beginning and continuous adaptations will be needed as new situations, either society- or pathogen-related, appear on scene. In these cases, direct transfer may be limited and changes in the original plan may be necessary. However, as the arrival of SARS-CoV-2 has shown, the urgent needs in fighting a new pathogen may restrict a clean user-centric design process. Consequently, we have to anticipate needs such that a good pandemic preparedness level can be achieved in advance. With the initial user feedback, we have shown that our tool was found to be intuitive and easy to use in its current stage. The iterative feedback loops with users and experts from local health authorities will further improve it and make it an important asset to fight future epidemics and pandemics.

\section*{Acknowledgements}

This work was performed as part of the project ‘Integrated Early Warning System for Local Recognition, Prevention, and Control for Epidemic Outbreaks’ (LOKI), which is funded by the Initiative and Networking Fund of the Helmholtz Association (grant agreement number KA1-Co-08). The authors PKB, JG, MZ, and MJK were also partially funded by the German Federal Ministry for Digital and Transport under grant agreement FKZ19F2211A.




\bibliographystyle{abbrv-doi}

\bibliography{
bibliography.bbl
}
\end{document}